\documentclass[showpacs,twocolumn,prl]{revtex4}

\usepackage{amssymb}

\usepackage{graphicx}

\begin{document}

\title{Revisiting numerical real-space renormalization group for quantum lattice
systems}

\author{Li-Xiang Cen}
\email{lixiangcen@scu.edu.cn}
\affiliation{Center of Theoretical Physics, College of Physical Science and
Technology, Sichuan University, Chengdu 610065, China}

\begin{abstract}
Although substantial progress has been achieved in solving quantum impurity
problems, the numerical renormalization group (NRG) method generally performs
poorly when applied to quantum lattice systems in a real-space blocking
form. The approach was thought to be unpromising for most lattice systems owing to
its flaw in dealing with the boundaries of the block.
Here the discovery
of intrinsic prescriptions to cure interblock interactions is reported which
clears up the boundary obstacle and is expected to reopen the application of
NRG to quantum lattice systems. While the resulting RG
transformation turns out to be strict in the thermodynamic limit, benchmark
tests of the algorithm on a one-dimensional Heisenberg antiferromagnet and a
two-dimensional tight-binding model demonstrate its numerical
efficiency in resolving low-energy spectra for the lattice systems.
\end{abstract}

\pacs{05.10.Cc, 05.30.Fk, 75.10.Jm}

\maketitle

Since the success in solving the Kondo problem by Wilson \cite{wilson},
there had been lots of attempts in applying the numerical
renormalization group (NRG) method to treat other quantum many-body problems
in a similar way. However, the numerical RG algorithm based on real-space
schemes met its Waterloo as it performed very poorly in subsequent several
applications. In particular, the approach was shown to give inaccurate results for
most interacting lattice systems \cite{lattice1,lattice2,lattice3}
and for the Anderson localization problem \cite{lee1,lee2}. The
difficulty, identified by White and Noack \cite{white1} via a simple
one-dimensional tight-binding model, lies in that the approach is flawed
in its treatment of boundaries of a block.

The inability to apply the NRG method to quantum lattice systems
is a heavy loss to the research of condensed matter physics.
Although a
closely related method, the density matrix RG algorithm \cite{white2,white3,whiter},
was proposed later and has been shown very effective in achieving
the ground-state energy for interacting lattice systems, there exist strong
restrictions of this method in its application to excitation spectra or to
systems with high spatial dimensions. The aim of this letter is to show that
the flaw of the NRG method exposed previously can be eliminated via
intrinsic prescriptions to cure interblock interactions and the derived
``regularized" real-space blocking version of the NRG scheme is able to yield
reliable results for quantum lattice systems.

In the standard real-space version of the NRG approach, one starts from
a block Hamiltonian
$H_L$ of $L$ sites and diagonalizes it exactly. By keeping a certain amount of
the lowest-energy states $\{|\psi _j\rangle ,j=1,\cdots ,m\}$, one then uses
them to construct a Hamiltonian for a larger system composed of two such
blocks: $H_{2L}=H_L^{[1]}+v_{12}+H_L^{[2]}$, where $v_{12}$ denotes the
interblock coupling. The primitive algorithm through projecting $H_{2L}$
simply onto $m^2$ tensor product states $|\psi _j\rangle \otimes
|\psi _{j^{\prime }}\rangle $ fails to achieve reliable results. The new
version of the NRG scheme here adopts a slightly different route: one should start by
diagonalizing a pair of block Hamiltonians, e.g., the one $H_L$ with an open
boundary condition and the one with a periodic form, $H_L^p$. Let us denote
correspondingly the low-lying eigenstates of $H_L^p$ by $\{|\psi _j^p\rangle \}$.
The effective performance of the NRG algorithm resides in that, instead of
using the set of tensor product states, the new pair of compound Hamiltonians,
the open $H_{2L}$ and the periodic $H_{2L}^p=H_{2L}+v_{21}$, are constructed and
diagonalized in virtue of the following two sets of $m^2$ states
\begin{equation}
|\Psi _{jj^{\prime }}\rangle =U_{adj}:|\psi _j\rangle \otimes |\psi
_{j^{\prime }}\rangle ,  \label{prsc1}
\end{equation}
and
\begin{equation}
|\Psi _{jj^{\prime }}^p\rangle =U_{adj}^{\otimes 2}:|\psi _j\rangle \otimes
|\psi _{j^{\prime }}\rangle ,  \label{prsc2}
\end{equation}
respectively. The transformation $U_{adj}$ here is the key prescription
introduced to cure the interblock coupling. It is formally a range-$L$
operator acting on adjacent segments of the compound block, namely,
on the sites from $\frac 12L+1$ to $\frac 32L$ to
generate $|\Psi _{jj^{\prime }}\rangle $, and on the regions of the above
sites and the sites from $\frac 32L+1$ to $\frac 12L$ separately so as to
generate $|\Psi_{jj^{\prime }}^p\rangle $.
As will be elucidated in detail, $U_{adj}$ involves two different expressions,
$U_{adj}^{(1)}$ and $U_{adj}^{(2)}$; the latter takes the form of
$U_{adj}^{(2)}=\sum_j|\psi _j^p\rangle \langle \psi _j^{\circlearrowleft }|$,
in which
$|\psi_j^{\circlearrowleft }\rangle \equiv \hat{\mathcal{T}}^{L/2}|\psi _j\rangle $
denotes a state through translating $|\psi _j\rangle $ by $L/2$
sites and $\hat{\mathcal{T}}|i_L\rangle =|i_1\rangle $ has been assumed.

The theoretical foundation of the above formulated prescription
rests on the following analysis about the utmost case in which the size
$L$ of the block is sufficiently large so
that the system $H_L$ can be viewed as thermally extensive.
By dividing the block into two subsystems with intermediate coupling,
$H_L=h_l+v_{lr}+h_r$, one can make an assertion that the low-lying
spectra of $H_L$ and of the disconnected Hamiltonian $h_l+h_r$ are identical:
$E_n=E_{n_1}^l+E_{n_2}^r$, where $E_{n_1}^l$ and $E_{n_2}^r$ denote the
spectra of the left half $h_l$ and of the right half $h_r$, respectively.
This is simply understood since the partition function of the block,
$\mathbf{Z}\equiv \mathrm{Tr}e^{-\beta H_L}$, should fulfill $\mathbf{Z}=
\mathbf{Z}_l\mathbf{Z}_r$ for any finite temperature $1/\beta $; the latter
identity, in which $\mathbf{Z}_{l,r}= \mathrm{Tr}e^{-\beta h_{l,r}}$,
actually accounts for the additivity of thermodynamic potentials of
macroscopic thermal systems.
As $H_L$ and $h_l+h_r$ possess the same low-lying spectra,
the low-energy portion of the two systems could be linked
by an isometry map, which turns out to delineate
the first form of the prescribed transformation: $U_{adj}^{(1)}=\sum_n|\psi
_n\rangle\langle \varphi _n|$, with $|\varphi_n\rangle \equiv
|\varphi_{n_1n_2}\rangle$ denoting the low-lying eigenstates of $h_l+h_r$.
That is, the low-energy portion of $H_L$ could be obtained via
$U_{adj}^{(1)}(h_l+h_r)[U_{adj}^{(1)}]^{\dagger}$, and the effect
of the intermediate coupling $v_{lr}$ hence is equally described by the
transformation $U_{adj}^{(1)}$.

It is then crucial to note that the formally range-$L$ $U_{adj}^{(1)}$
indeed has an effective range less than $L$ as the block size $L$ increases
asymptotically. Especially, for the infinite $H_L$ with a gapped ground state,
the range of $U_{adj}^{(1)}$ turns out to be local since the evolution operator
generated by $H_L$ in a finite period $\tau$, formed as
$e^{-ih_l\tau}\otimes e^{-ih_r\tau}$
sandwiched by $U_{adj}^{(1)}$ and $[U_{adj}^{(1)}]^{\dagger}$,
could be simulated by a quantum circuit \cite{QIT} with finite
depth. The latter representative, which describes the time
evolution operator in virtue of a few layers of piecewise local unitary
operators, has also been exploited to characterize quantum phases
and topological orders for quantum many-body systems \cite{wen1,wen2}.

Consequently, one can employ $U_{adj}^{(1)}$ further
to connect the pair of Hamiltonians $H_L^p$ and
$H_L^{\circlearrowleft }\equiv H_L^p-v_{lr}$, in which
$H_L^{\circlearrowleft }$, with edges inside the block,
is simply a Hamiltonian achieved via translating $H_L$ by $L/2$ sites.
An additional condition to simulate the effect of $v_{lr}$ here
is that the effective range of $U_{adj}^{(1)}$ should be less than $L$,
i.e., $U_{adj}^{(1)}$ should commute with $v_{rl}$.
The second form of $U_{adj}$,
$U_{adj}^{(2)}=\sum_j|\psi _j^p\rangle \langle \psi _j^{\circlearrowleft }|$,
is then yielded
and the contained $|\psi _j^{\circlearrowleft }\rangle $ are just
eigenstates of $H_L^{\circlearrowleft }$. By the same token, as the
compound systems $H_{2L}$ and $H_{2L}^p$ are considered,
the effects of interblock couplings $v_{12}$ and $v_{21}$ on their
low-energy behavior could be simulated by imposing the transformations,
either $U_{adj}^{(1)}$ or $U_{adj}^{(2)}$, on corresponding adjacent
segments of the disconnected system $H_L^{[1]}+H_L^{[2]}$.

The analysis above reveals that the regularized basis sets of equations (\ref{prsc1})
and (\ref{prsc2}), or more accustomedly, their symmetric and anti-symmetric
combinations
$|\Psi_{J\pm}\rangle \equiv |\Psi _{jj^{\prime }}\rangle \pm |\Psi _{j^{\prime }j}\rangle $
and
$|\Psi_{J\pm}^p\rangle \equiv |\Psi _{jj^{\prime }}^p\rangle \pm |\Psi _{j^{\prime }j}^p\rangle $,
specify exactly the low-energy solutions for compound blocks $H_{2L}$ and
$H_{2L}^p$ as $L\rightarrow \infty $. It thus illuminates a regularized
version of the NRG scheme
by incorporating the described transformation into the algorithm
in order to reconstruct Hamiltonians for doubly increasing blocks.
Historically, the speculation through applying
a variety of boundary conditions to perform the NRG procedure has
ever been endeavored by White and Noack in dealing with the tight-binding
model and the localization problem of one dimension \cite{white1,white4}.
The discovery of the distinct prescription here should enable us to exploit
the NRG method for general quantum lattice systems, i.e., for
interacting systems and for systems with high spatial dimensions.

The key of the present algorithm rests on recognizing correspondences among
basis states of the Hamiltonians with different boundary configurations
so as to build the transformation $U_{adj}$. For this purpose, one should
separate the basis states by quantum numbers and then discern the matching of
states among different sets but with identical quantum numbers through
comparing their fidelities. Due to finite-size
effects, the matching of states would become less
evident as the energy level increases; it requests consequently that the
initial block to be exactly diagonalized should be of considerable size.
It should be noted that, in spite of the correspondences having been
established, $U_{adj}$'s are still not uniquely determined owing to the
phase uncertainty of the eigenstates. This uncertainty, however, doesn't
affect the basic efficiency of the scheme since different choices of
the phases have no influence on the transformations under $L\rightarrow \infty$.
In practical calculations it can be simply removed, e.g., by setting
$\langle \psi _k|\varphi _k\rangle $ (or $\langle \psi
_k^\circlearrowleft|\psi _k^p\rangle $) to be real and positive \cite{note}.

\begin{table}[b]
\caption{Energies of $21$ low-lying states of the $L=16$ Heisenberg chain
obtained by the regularized
NRG scheme using $U_{adj}^{(1)}$. Presented also includes the information
of the degeneracy $f$, quantum numbers of the total spin $s$ and the momentum
$k$ of the corresponding energy levels yielded by the symmetry-preserved algorithm.
An accuracy about $10^{-5}$ is obtained by keeping $m=47$ states
(cf. the values of the ground and the first-excited energies achieved by exact
diagonalization $E_0\approx-22.44681$ and $E_1\approx-22.00401$ \cite{exact1,exact2}).}
\label{heisen}
\begin{ruledtabular}
\begin{tabular}{cccc}
~~      &   $s;k$   & $m=21$ &  $m=47$ \\ \hline
$E_0~(f=1)$	&	$s=0;k=0$	 &	-22.44170	&	-22.44639 	 \\
$E_1~(f=3)$	&	$s=1;k=8$	 &	-21.99120	&	-22.00326 	 \\
$E_2~(f=6)$	&	$s=1;k=7,9$	 &	-21.36381	&	-21.37456 	 \\
$E_3~(f=5)$	&	$s=2;k=0$	 &	-21.27190	&	-21.28826 	 \\
$E_4~(f=6)$	&	$s=1;k=1,15$ &	-21.06151	&	-21.07610 	 \\
\end{tabular}
\end{ruledtabular}
\end{table}

To verify the efficiency of this new version of the NRG scheme, the algorithm
has been implemented to calculate a periodic spin-1 antiferromagnetic
Heisenberg model $H=\sum_i\mathbf{S}_i\cdot \mathbf{S}_{i+1}$ with $16$
sites. The initial $H_L$ and $h_l+h_r$ with $L=8$ are exactly
diagonalized and the transformation $U_{adj}^{(1)}$ is employed to generate
the basis states for $H_{2L}^p$. The algorithm recovers the exact
energy \cite{exact1,exact2} to at least $5$
digits (with $m=47$ or more states to be kept). As the derived low-lying spectra
are shown in Table I, some key points to perform the algorithm are recited as
below. (1) It is convenient to invoke four-component tensors to
represent the states: $|\psi _j\rangle \otimes |\psi _{j^{\prime }}\rangle
\equiv \psi _{j_lj_rj_l^{\prime }j_r^{\prime }}$, where the indices $j_l$,
$j_r$, $j_l^{\prime }$ and $j_r^{\prime }$ account for the four half blocks with
size $L=4$, respectively. Transformation of $U_{adj}$ on adjacent halves is
then realized by summations over indices $(j_r,j_l^{\prime })$ or
$(j_r^{\prime },j_l)$. (2) Symmetries related to the total spin $S^2$ and $S_z
$ can be exploited to reduce the computing cost since these invariants are
preserved under the transformation $U_{adj}$. States with definite quantum
numbers $(s,s_z)$ for the $L=16$ block are readily constructed by those of
$L=8$ in virtue of Clebsch-Gordan coefficients \cite{bohm}, which are then
acted on by the transformation
$U_{adj}^{\otimes 2}$ and are Gram-Schmidt orthonormalized afterwards
so as to get the desired basis states,
$|\Psi _{J,s,s_z}^p\rangle $. (3) The algorithm described till now does not
involve the translational symmetry except for that of the blocked translation
with $8$ sites. To cope with the full translational symmetry, one should
construct states
with definite momentum quantum number $k$ using the basis states
$|\Psi _{J,s,s_z}^p\rangle $ and
$\hat{\mathcal{T}}^i|\Psi _{J,s,s_z}^p\rangle $ $(i=1,\cdots ,7)$, and
then diagonalize the Hamiltonian $H_{2L}^p$ in symmetry-preserved spaces with
quantum numbers $(s,s_z,k)$.

Extensions of the proposed NRG scheme to lattices with more than
one spatial dimension
could be naturally procured. Take the square lattice as an example:
by sketching out four different boundary configurations of an
$L\times L$ lattice [see Fig. 1 (a) and (b)], the transformations $U_{adj}^{(1)}$
and $U_{adj}^{(2)}$ could be constructed from eigenstates of
pair configurations, either (i) and (ii) of Fig. 1 (a) or
(iii) and (iv) of Fig. 1 (b), respectively.
Since the lattice Hamiltonian usually possesses symmetries related to a
specific crystallographic point group, classifying degenerate
eigenstates via irreducible group representations \cite{chen} should be a task
prerequisite to the procedure.
The compound system of the quadrupled block needs to be represented by a
16-component tensor with indices illustrated in Fig. 1 (c). Basis
states for the block Hamiltonian with periodic boundary conditions
are obtained by imposing $U_{adj}$ on each joining region
with fourfold adjacent edges, which can be realized via
summations over corresponding indices, respectively.

\begin{figure}[t]
\includegraphics[width=0.8\columnwidth]{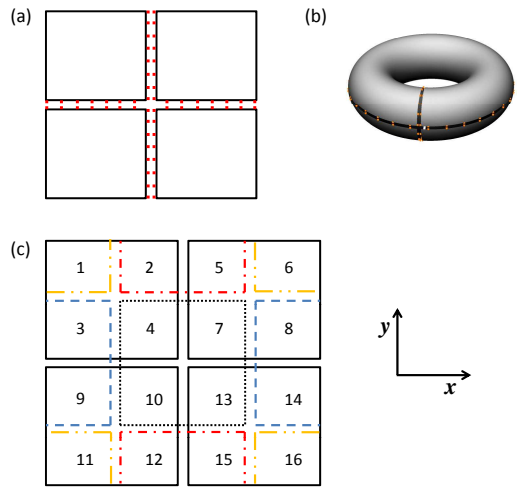}
\caption{Schematic of configurations of the square lattice.
(a) Configurations (i) and (ii) of the $L\times L$ lattice with
open boundaries, in which inside edges are either disconnected [configuration
(i)] or connected [configuration (ii)]. (b) Lattice configurations (iii)
and (iv) with connected boundaries. The one with disconnected inner edges
[configuration (iii)] is equivalent to the configuration (ii) up to
a translation along both $x$ and $y$ directions by $L/2$ sites.
(c) Configuration of the representative tensor for the compound
quadrupled lattice. The joining regions of adjacent segments are indicated by
indices $\{4,7,10,13\}$, $\{8,3,14,9\}$, $\{12,15,2,5\}$, and $\{16,11,6,1\}$,
on which $U_{adj}$ should be imposed separately.}
\end{figure}

As a test case, the two-dimensional scheme is examined by a tight-binding
model, a single particle hopping on a square lattice with a Hamiltonian
\begin{equation}
H=\sum_{x,y}[(ua_{x+1,y}^{\dagger }a_{x,y}+h.c.)+(va_{x,y+1}^{\dagger
}a_{x,y}+h.c.)],
\end{equation}
where $a_{x,y}^{\dagger }$ and $a_{x,y}$ are the creation and annihilation
operators on the site $(x,y)$ and the hopping parameters are set as $u=-1$
and $v=-3/2$. The Hamiltonians of an $L=64$ square lattice with various boundary
configurations are exactly diagonalized of which the eigenstates of
configurations (ii) and (iv) are recorded as
$|\psi _j\rangle$ and $|\psi _j^p\rangle$, respectively.
Unitary $U_{adj}$'s can be built for the present system through full
correspondences of the derived bases, which enables us to calculate states
for the quadrupled lattice at any energy level. Starting from $m$ states
$\{|\psi _j\rangle \}$ closest to a particular energy, the $4m$ basis states
of the compound periodic lattice
are generated and divided into four sets, $\{|\Psi _j^{\pm ,\pm }\rangle
,j=1,\cdots ,m\}$, in which $``\pm "$ denotes the symmetric or anti-symmetric
property of states under $\hat{t}_{x(y)}\equiv \hat{\mathcal{T}}_{x(y)}^L$,
translations along the $x$ $(y)$ direction by $L$ sites.
The lowest energies of the $L=128$ system achieved by the algorithm
via $U_{adj}^{(1)}$ are shown in Table II,
with which the exact energies and the energies achieved by the primitive NRG
are presented for reference.

\begin{table*}[t]
\caption{Lowest energies of the $L=128$ square lattice with periodic
boundary conditions achieved by the regularized NRG algorithm via
the prescription $U_{adj}^{(1)}$. The energies are calculated in separated
spaces of the four sets of basis states $\{|\Psi_j^{\pm,\pm}\rangle,j=1,\cdots,m\}$
with $m=64$ and are listed according to their values.
The results recover the exact energies to $4$-$5$ digits and exhibit
correct order, while the results obtained
by the primitive NRG scheme are quite inaccurate and fail to yield
the same ordering.}
\label{twod}
\begin{ruledtabular}
\begin{tabular}{ccccc|ccccc}
~~ & $\{\hat{t}_x,\hat{t}_y\}$ & Exact & Regularized RG & Primitive RG & ~~ & $\{\hat{t}_x,\hat{t}_y\}$ & Exact & Regularized RG & Primitive RG \\ \hline
$E_0$	&	$	\{+,+\}	$	&	-5.00000	&	-4.99960 	&	-4.99462	 &  $E_8$	&	$	\{-,-\}	$	&	-4.99398	 &	-4.99386	 &	-4.97853	    \\
$E_1$	&	$	\{-,+\}	$	&	-4.99759	&	-4.99732 	&	-4.99431	 &  $E_9$	&	$	\{+,+\}	$	&	-4.99037	 &	-4.99011 	 &	-4.98698	    \\
$E_2$	&	$	\{-,+\}	$	&	-4.99759	&	-4.99728 	&	-4.98821	 &	$E_{10}$  &	$	\{+,+\}	$	&	-4.99037	 &	 -4.99009 	 &	-4.98314	 	\\
$E_3$	&	$	\{+,-\}	$	&	-4.99639	&	-4.99623 	&	-4.99417	 &	$E_{11}$  &	$	\{+,-\}	$	&	-4.98676	 &	 -4.98674 	 &	-4.98494	 	\\
$E_4$	&	$	\{+,-\}	$	&	-4.99639	&	-4.99617 	&	-4.98653	 &	$E_{12}$  &	$	\{+,-\}	$	&	-4.98676	 &	 -4.98672 	 &	-4.97730	 	\\
$E_5$	&	$	\{-,-\}	$	&	-4.99398	&	-4.99395 	&	-4.99387	 &	$E_{13}$  &	$	\{+,-\}	$	&	-4.98676	 &	 -4.98668 	 &	-4.97698	 	\\
$E_6$	&	$	\{-,-\}	$	&	-4.99398	&	-4.99391 	&	-4.98776	 &	$E_{14}$  &	$	\{+,-\}	$	&	-4.98676	 &	 -4.98666 	 &	-4.96774	 	\\
$E_7$	&	$	\{-,-\}	$	&	-4.99398	&	-4.99389 	&	-4.98463	 &	$E_{15}$  &	$	\{+,+\}	$	&	-4.98555	 &	 -4.98541 	 &	-4.97742	 	\\
\end{tabular}
\end{ruledtabular}
\end{table*}

Performance of the algorithm employing the prescription $U_{adj}^{(2)}$
on the above two models,
the spin-$1$ Heisenberg chain and the tight-binding model, displays that
it is slightly less accurate than that utilizing $U_{adj}^{(1)}$.
For example, the algorithm using $U_{adj}^{(2)}$ yields a ground-state
energy $E_0\approx -22.43222$ for the 16-site Heisenberg chain
provided that the same amount of $47$ states are kept.
Nevertheless, it is of interest to mention
an exceptional case of the tight-binding model,
where the numerical calculation discloses that the symmetric basis states
$\{|\Psi _j^{+,+}\rangle \}$,
generated via imposing $U_{adj}^{(2)}$'s
on the state vectors of direct sum
$|\psi _j\rangle \oplus |\psi _j\rangle \oplus |\psi _j\rangle \oplus |\psi
_j\rangle $, are exact eigenstates of the quadrupled system
with periodic boundary conditions.
In view that the intercepted state vector of any joining segments with
fourfold adjacent edges represents a basis state
of the lattice configuration $(iii)$, the transformation $U_{adj}^{(2)}$
on each of those joining segments maps it precisely onto $|\psi_j^p\rangle $,
a basis state of the lattice configuration (iv).
That is to say, direct combinations of basis states of the periodic
$L\times L$ block,
$|\psi _j^p\rangle \oplus |\psi _j^p\rangle \oplus |\psi _j^p\rangle \oplus |\psi
_j^p\rangle $, already give rise to one fourth of the full set
of exact solutions of the compound $2L\times 2L$ system.
This result is indeed general: it is independent of
the size $L$ and applies also to the one-dimensional tight-binding model
wherein it is very easy to verify.

In the present algorithm at least two truncated
basis sets are kept and they should be expressed in a complete set of
spin bases. This should be the case at each iteration since
any such truncated set is too incomplete to represent other basis
states with different boundary configurations. As a result, repetition of the
described RG procedure is constrained by the capacity of storing vectors with
exponentially growing dimensions.
In this sense, the regularized NRG scheme is applicable to finite lattices
rather than the infinite one since the iterative performance of the procedure
for the latter system will break down inevitably.
This indeed offers the underlying cause which supports partly the folklore
stating that all real-space RG schemes
are necessarily inaccurate for lattice systems.

Alternatively, if one chooses to sacrifice some of the accuracy,
an iterative procedure could be achieved, at least for
one-dimensional systems, by projecting the derived basis states, e.g., those
of $H_{2L}$ and $H_{2L}^p$, on the tensor product space of the $m$
states kept to get bases $|\tilde{\psi}_j\rangle$ and
$|\tilde{\psi}_j^p\rangle$ of dimension $m^2$.
The prescribed transformation, $\tilde{U}_{adj}$, could be built by virtue of
these basis states which is then imposed on the states
$|\tilde{\psi}_j\rangle \otimes |\tilde{\psi}_{j^\prime }\rangle$ to
generate basis states for the pair of Hamiltonians with larger size.
The truncation of the algorithm here is distinctly different from that of the
primitive NRG scheme in view that the $m$ states kept are taken from a Hilbert
space of dimension $m^4$ instead of dimension $m^2$.
It is also worthy to note that the strategy adopted here, projecting
the basis states of the compound block onto a restricted space
of tensor product states of subsystems, is what has been done
in the contractor RG method \cite{core1,core2,core3}.
In comparison, the interblock coupling of the effective Hamiltonian
in the latter scheme is obtained by substracting the contributions
of contained subclusters, while in the present scheme the effective
Hamiltonian of the compound system could be constructed
directly by virtue of the regularized basis states.

The author thanks H.-G. Luo for helpful discussions.
Support of the National Natural Science Foundation
of China (Grant No. 10874254) is acknowledged.


\begin{references}
\bibitem{note} For topologically nontrivial systems the choice of the phases
might not be irrelevant and it should be investigated elsewhere.
\end{references}
\end{document}